\documentclass[a4paper,11pt]{article}
\usepackage{pos}
\usepackage[capitalize]{cleveref}

\newcommand{\SU}{\mathrm{SU}}

\title{Progress in Normalizing Flows for 4d Gauge Theories}

\author*[a,b]{Ryan~Abbott}
\author[a,b]{Denis~Boyda}
\author[c,a,b]{Daniel~C.~Hackett}
\author[d,e]{Gurtej~Kanwar}
\author[e,a,b]{Fernando~Romero-L\'opez}
\author[a,b]{Phiala~E.~Shanahan}
\author[a,b]{Julian~M.~Urban}

\affiliation[a]{Center for Theoretical Physics, Massachusetts Institute of Technology, Cambridge, MA 02139, USA}
\affiliation[b]{The NSF AI Institute for Artificial Intelligence and Fundamental Interactions}
\affiliation[c]{Fermi National Accelerator Laboratory, Batavia, IL 60510, U.S.A.}
\affiliation[d]{Higgs Centre for Theoretical Physics, University of Edinburgh, Edinburgh EH9 3FD, UK}
\affiliation[e]{Albert Einstein Center, Institute for Theoretical Physics, University of Bern, 3012 Bern, Switzerland}

\abstract{Normalizing flows have arisen as a tool to accelerate Monte Carlo sampling for lattice field theories. This work reviews recent progress in applying normalizing flows to 4-dimensional nonabelian gauge theories, focusing on two advancements: an architectural improvement referred to as learned active loops, and the application of correlated ensemble methods to QCD with $N_f=2$ dynamical fermions.}

\FullConference{The 41st International Symposium on Lattice Field Theory (LATTICE2024)\\
 28 July - 3 August 2024\\
Liverpool, UK\\}

\begin{document}
\maketitle

\section{Introduction}
Monte Carlo importance sampling forms the basis of all lattice field theory calculations. Many practical developments over the past decades---Hybrid Monte Carlo, multigrid preconditioners, and efficient numerical codes---have enabled precision studies of dynamical lattice gauge theories, including lattice QCD. Despite these advances, critical slowing down and topological freezing limit the lattice spacings accessible in state-of-the-art calculations, which has lead to recent interest in new algorithms, such as those provided by machine learning.
One such approach has been the application of normalizing flows to lattice field theory, which has been demonstrated to eliminate critical slowing down in certain lattice field theories~\cite{Albergo:2019eim,Kanwar:2020xzo,Albergo:2022qfi} and, for example, may provide near-term practical advantages through correlated sampling to access specific observables~\cite{Abbott:2024kfc,Bacchio:2023all}.
Normalizing flow methods have also been extended to Monte Carlo sampling for dynamical lattice QCD at a coarse lattice spacing~\cite{Abbott:2022hkm}. These promising early results have motivated recent efforts to improve the quality of models and find new practical applications of flows.

The present work is focused on two recent developments in applying normalizing flows to 4d gauge theories, particularly QCD. The first is an architectural improvement called ``learned active loops''. This method replaces what has typically been a static choice in previous models with a learned component, expanding the ability for each layer in the flow to modify the physical degrees of freedom and substantively improving model quality.
The second development is the application of the correlated ensemble methods of Ref.~\cite{Abbott:2024kfc} to a system with dynamical fermions, specifically to calculate the gluon momentum fraction of the pion in $N_f=2$ QCD with twisted mass fermions at $M_\pi \simeq 540$ MeV. This is the first application of this method to a theory with dynamical fermions. For this particular study, a computational advantage of the approach that incorporates flows is shown, even when training costs are taken into account.

\section{Learned Active Loops}
This work introduces learned active loops as an architectural improvement for the coupling-based spectral flow architectures described in Ref.~\cite{Abbott:2023thq}. Coupling-based flows are constructed by stacking several layers, each of which transforms some subset $\{(x,\mu)\}_A$ of the gauge links $U_\mu(x)$, referred to as the active links.
Transformations of these active links are defined by first constructing untraced Wilson loops containing each active link, then transforming these loops, then ``pushing back'' the transformation onto the active links. 
The constructed loops are referred to as \emph{active loops}. 
Spectral flows in particular use an eigenvalue decomposition of the active loops to isolate and act on gauge-invariant information~\cite{Abbott:2023thq,Kanwar:2020xzo,Boyda:2020hsi}.

Previous flow architectures have typically used a fixed active loop geometry
per layer, 
often using a particular orientation of a small loop, such
as an untraced plaquette or $2 \times 1$ loop, alternating the direction
across successive layers. 
The central idea behind learned active loops
is to replace this fixed choice with a learned component, adding
additional parameters that specify a linear combination of
loops to transform in each layer. This is a strict generalization of the previous approach, and expands the degrees of freedom acted on by each layer of the flow. 
Intuitively, the benefit of
this approach is that it allows the transformation to more directly access and correlate both small and large active loops in an individual layer, which combinatorially increases the space of transformations.

The implementation of learned active loops in this work has two components: first, the frozen degrees of freedom are passed through a ``staple network'' to produce an ``active staple'', denoted by $S^\text{act}_\mu(x)$. Here a staple $S_\mu(x)$ refers to any set of complex matrices derived from gauge links that transforms like the gauge link $U_\mu(x)$ with the same $(x,\mu)$ under gauge transformations (notably a staple is \emph{not} required to be an element of $\SU(N)$).
Once the active staple has been computed, the active loop is then defined by
\begin{equation}\label{eq:active-loop-def}
    L_\mu(x) = \mathcal{P}_{\SU(N)}(U_\mu(x) S^\text{act}_\mu(x)^\dag)
\end{equation}
where $\mathcal{P}_{\SU(N)}$ denotes a conjugation equivariant projection from general complex matrices onto $\SU(N)$.
Here, conjugation equivariant means that the projection must satisfy
\begin{equation}
\label{eq:conj-equiv}
\mathcal{P}_{\SU(N)}(g M g^{\dag}) = g \mathcal{P}_{\SU(N)}(M) g^{\dag}
\end{equation}
for any $g \in \SU(N)$.

In the architectures of this work, each coupling layer contains its own independent staple network as a subcomponent.
Each such staple network is
constructed out of a series of $N_\ell$ layers, which successively transforms the input staples as
\begin{equation}
    S^{(0)}_{\mu k} \to S^{(1)}_{\mu k} \to \dots \to S^{(N_\ell)}_{\mu k}.
\end{equation}
The index $k$ is an additional channel index that ranges from $1$ to $N_h^{(i)}$. All staple networks used in this work fix $N_h^{(0)} = N_h^{(N_\ell)} = 1$ and $N_h^{(i)} = N_h$; the constant $N_h$ is referred to as the width of the network.
The layers involved in this work are all linear transformations on the space of all staples, though any gauge-equivariant architecture could be used here. For example, the L-CNN architectures from Ref.~\cite{Favoni:2020reg} could be explored as an alternative architecture for the staple network.

The inputs to the network are initialized by setting the staples equal to the frozen links, zeroing out any active links by setting
\begin{equation}
S_{\mu 0}^{(0)}(x) = (1 - m_{\mu}(x)) U_{\mu}(x),
\end{equation}
where $m_{\mu}(x)$ denotes the active mask, defined by
\begin{equation}
m_{\mu}(x) = 
\begin{cases}
1 & U_{\mu}(x) \text{ is active} \\
0 & U_{\mu}(x) \text{ is frozen}.
\end{cases}
\end{equation}
Schematically, each successive layer acts by parallel transporting same-direction staples from neighboring sites and then taking a linear combination.
In detail, to apply successive layers in the staple network, first left and right staples are constructed out of the input staples via
\begin{equation}
\begin{aligned}
    S^{(i)L}_{\mu\nu k} &= S^{(i)\dag}_\nu(x + \hat{\mu} - \hat{\nu}) S^{(i)\dag}_\mu(x - \hat{\nu}) S^{(i)}_\nu(x - \hat{\nu}), \\
    S^{(i)R}_{\mu\nu k} &= S^{(i)}_\nu(x + \hat{\mu}) S^{(i)\dag}_\mu(x + \hat{\nu}) S^{(i)\dag}_\nu(x).
\end{aligned}
\end{equation}
These new staples are then concatenated and flattened along with the original staples $S^{(i)}_{\mu k}$, forming an augmented vector $S^{(i),\text{aug}}_{\mu k}$---i.e., the $\nu$ and $k$ indices of the three original objects $S^{(i)}_{\mu\nu k}$ and $S^{(i)L/R}_{\mu\nu k}$ are subsumed into the $k$ index of $S^{(i),\text{aug}}_{\mu k}$. Finally, the different elements $k$ of the augmented vector are linearly combined to form the outputs of the layer via
\begin{equation}
    S^{(i+1)}_{\mu k}(x) = \sum_{\ell} c^{(i)}_{k \ell} S^{(i),\text{aug}}_{\mu \ell}(x),
\end{equation}
where $c^{(i)}_{k \ell}$ are learned coefficients.
After iterating through the layers of the staple net, the active staple is taken to be the sole element of the final set of staples:
\begin{equation}
    S^\text{act}_\mu(x) = S^{(N_\ell)}_{\mu 0}(x).
\end{equation}

After the active staple is chosen, what remains is to define the conjugation-equivariant projection $\mathcal{P}_{\SU(N)}$ used in \cref{eq:active-loop-def}. 
In order to satisfy \cref{eq:conj-equiv}, the flow models in this work utilize a
two-step projection. First the matrix is projected onto $U(N)$ via
polar projection, defined by
\begin{equation}
\label{eq:polar-proj}
\mathcal{P}_{U(N)}(M) = (M^{\dag} M)^{-1/2} M.
\end{equation}
In practice this projection can be implemented by performing a
singular value decomposition $M = U \Sigma V^{\dag}$, and returning
$\mathcal{P}_{U(N)}(M) = U V^{\dag}$, which is equivalent to, but more
numerically stable than, \cref{eq:polar-proj}.
Once the polar projection is computed, the matrix can be projected
onto $\mathrm{SU}(N)$ by dividing out the determinant:
\begin{equation}
\label{eq:proj-suN}
\mathcal{P}_{\mathrm{SU}(N)}(M) = \frac{\mathcal{P}_{U(N)}(M)}
{(\det [\mathcal{P}_{U(N)}(M)])^{1/3}}.
\end{equation}
It is straightforward to check via \cref{eq:polar-proj,eq:proj-suN} that this projection satisfies \cref{eq:conj-equiv}, and hence is conjugation-equivariant.

To test the effectiveness of learned active loops, we perform a simple benchmark using two models, with and without learned active loops.
Both models are spectral flow models constructed similarly to the models of Ref.~\cite{Abbott:2023thq}, combining direction and location updates, repeated a total of 3 times for $3 \times (48 + 16) = 192$ layers. The only difference between the two models is that one model uses learned active loops in its direction updates while the other uses a fixed active loop, namely the untraced plaquette.
Both models are trained using the Adam optimizer~\cite{kingma2017adam} with path gradients~\cite{vaitl2022gradients}.
For the model with learned active loops, the staple networks are constructed using 2 hidden layers with $N_h = 8$, and the coefficients of the staple network are initialized using the PyTorch $\texttt{xavier\_normal\_}$ initialization scheme~\cite{glorot2010understanding} with a gain of $0.5$. In addition, the weights $c^{(i)}_{k\ell}$ of the staple network are initialized near the identity, meaning that the weights are set so that initialization with a gain of $0$ would result in each layer of the staple network producing the identity function.

\begin{figure}
\begin{center}
    \includegraphics[width=10cm]{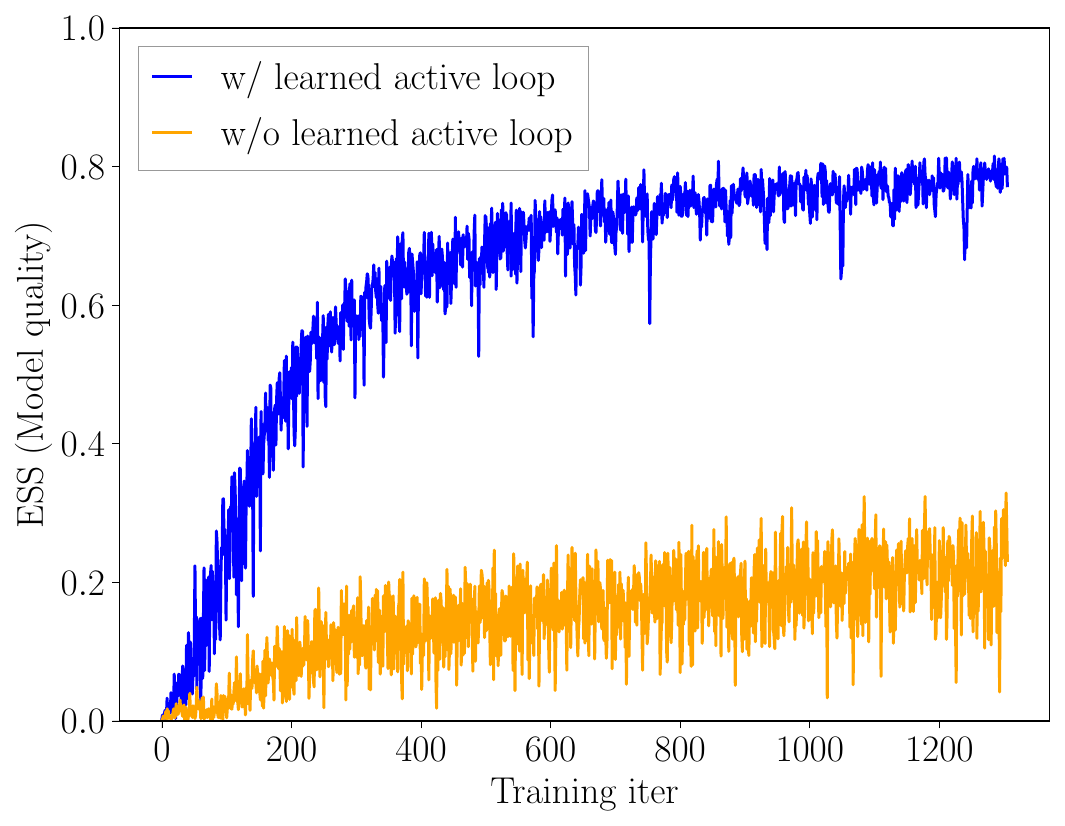}
\end{center}
    \caption{\label{fig:learned-active-loop}Example training curves with and without learned active loops. The models are otherwise identical, and the target theory is a pure-gauge plaquette action with $\beta = 2$ on a $4^4$ lattice, optimized as described in the text.}
\end{figure}

The results of training both models to target a pure gauge theory with a plaquette action at coupling $\beta=2$ on a $4^4$ lattice is shown in \cref{fig:learned-active-loop}. The model quality here is demonstrated by the effective sample size (ESS), defined by~\cite{doucet2001sequential,liu2001monte}
\begin{equation}
  \mathrm{ESS} = \frac{\mathbb{E}_{U \sim q(U)}[w(U)]^2}{\mathbb{E}_{U
	\sim q(U)}[w(U)^2]},
\end{equation}
where $w(U)=e^{-S(U)}/q(U)$ is the (unnormalized) reweighting factor, $S(U)$ is the target action defining the target density $p(U) \propto e^{-S(U)}$, and $q(U)$ is the model density.
The ESS can be thought of as the effective number of independent samples per model sample, with an ESS of 1 indicating a perfect model. The model that uses learned active loops shows a clear advantage over the model without, achieving almost $80\%$ ESS in comparison to $20\%$. Similar results have also been seen across a wide variety of different tests, indicating that learned active loops appear to be a strict improvement over previous spectral flow models.

\section{Correlated ensembles for hadron structure in $N_f=2$ QCD}

One promising application of flow models is as a map between two nearby distributions with a small difference in action parameters, $\Delta \lambda$. Such an application of flow models, extensively discussed in Ref.~\cite{Abbott:2024kfc}, can be used to compute derivatives with respect to action parameters:
\begin{equation}
    \frac{d \langle  O \rangle}{d \lambda} \simeq \frac{\langle  O \rangle_{\lambda_1} - \langle  O \rangle_{\lambda_2}}{ \lambda_1 - \lambda_2},
\end{equation}
where the right-hand side is a finite-difference approximation of the derivative, and $\langle \mathcal O \rangle_{\lambda}$ indicates the expectation value in the theory with action parameter $\lambda$. Using a flow model, this derivative can be computed as
\begin{equation}
   \frac{d \langle  O \rangle}{d \lambda} \simeq \big\langle {O}(U)-w(f(U)) \, {O}(f(U)) \big\rangle_{\lambda_1},
\end{equation}
where $f(U)$ is the flow acting on the gauge fields, and $w(f(U))=p_{\lambda_2}(f (U ))/q(f (U ))$ is the corresponding reweighting factor for the flow from $\lambda_1$ to $\lambda_2$. The same derivatives can be computed also by other means, such as $\epsilon$-reweighting,  or using independent ensembles generated at $\lambda_1$ and $\lambda_2$. The methods that incorporate flows were shown in Ref.~\cite{Abbott:2024kfc} to offer reduced uncertainties at fixed number of configurations, due to the possibility of using larger $\Delta \lambda$ while benefiting from correlated cancellations.

A particularly promising application of this idea is to the computation of hadronic matrix elements using Feynman-Hellmann techniques. The main idea is to add a term to the QCD action:
\begin{equation}
     S = S_{\rm QCD} + \lambda \mathcal O, 
\end{equation}
where $\mathcal O$ is some operator of interest. Then, the matrix element of that operator between hadronic states can be computed as a derivative of the hadron mass, $M_h$:
\begin{equation}
    \langle h | O | h \rangle =
    \left.\frac{1}{2 M_h} \frac{d M_h}{d \lambda}\right|_{\lambda \rightarrow 0}.
\end{equation}
 One example is an operator related to the gluon momentum fraction of the pion:
\begin{equation}
    \mathcal{O}=-\frac{\beta}{N_c} \operatorname{Tr} \operatorname{Re}\left(\sum_i U_{i 0}-\sum_{i<j} U_{i j}\right).
\end{equation}
This operator can be added as an asymmetry between the coupling of  spatial and temporal plaquettes, $U_{ij}$ and $U_{i0}$, respectively. The gluon momentum fraction can then be computed as
\begin{equation}
    \left.\frac{d M_h}{d \lambda}\right|_{\lambda \rightarrow 0}=-\frac{3 M_h}{2}\langle x\rangle_g^{\mathrm{latt}} \simeq \frac{M_h(\lambda)- M_h(0)}{\lambda}.
\end{equation}

In Ref.~\cite{Abbott:2024kfc}, we presented a proof-of-concept application of this approach in quenched QCD with a small lattice volume and quark masses corresponding to a larger-than-physical pion mass, and observed a >10$\times$ reduction of variance with a fixed number of configurations. Given this promising result, the next step is to perform the same calculation in a more physical setting with dynamical fermions, a larger lattice volume, and at lower pion mass. Below, we present our update on that front. In particular, we use $N_f=2$ twisted mass fermions using the L\"uscher-Weisz gauge action. Specifically, the action parameters and geometry are
\begin{equation}
    \kappa = 0.164111, \quad  \beta=3.8, \quad \mu=0.012, \quad L^3\times T = 12^3 \times 24.
\end{equation}
This follows the same setup as the ``A'' $N_f=2$ ensembles by the ETMC collaboration, see Ref.~\cite{Urbach:2007rt}. The pion mass is $a M_\pi = 0.276(2)$. Using the lattice spacing $a=0.1$ fm quoted in Ref.~\cite{Urbach:2007rt}, this corresponds to $M_\pi \simeq 540 $ MeV. The volume is $M_\pi L \sim 3.31$. We use 10k configurations generated with Chroma~\cite{Edwards:2004sx}. For the Feynman-Hellmann calculation, we use $\lambda=0.005$.

Since this is dynamical QCD, the flow approach has to be adapted to the presence of fermions. For this, we utilize the pseudofermion approach of Ref.~\cite{Abbott:2022zhs}. Specifically, we have a marginal model and a conditional model, which are trained separately. The marginal model acts on the gauge fields, while the conditional model is used to reduce the variance in the stochastic estimator of the fermion determinant ratio. For the marginal model, we use the same architecture presented in Ref.~\cite{Abbott:2024kfc}. For the conditional model we follow the approach of Ref.~\cite{Abbott:2022zhs}, with a significant modification: since both the base and target distributions are nontrivial, we allow for parallel transports with the flowed and unflowed fields. Further details will be presented in an upcoming publication.

The marginal model is trained using the action with an explicit fermion determinant at a small volume, $V=4^4$. It has 48 layers alternating the M2, M4, M2 and M4 masking pattern described in Ref.~\cite{Abbott:2024kfc}. 
It achieves 99.7\% ESS, which is an increase over the 94\% ESS achieved with an untrained identity flow. The conditional model has 16 layers with an alternating masking pattern and is also trained at $V=4^4$. It uses even-odd preconditioning, and it 
achieves a 99.6\% ESS, higher than the baseline 99.4\% ESS when using the identity pseudofermion flow. Both models have been trained with path gradients. 
At the evaluation volume, $12^3 \times 24$, the ESS of the marginal model with no pseudofermion flow and four pseudofermion noise shots is 53\%. Including the trained conditional model, it raises to 58\%. Without a trained marginal model, i.e.~simply using unimproved reweighting, the ESS would be $<10^{-3}$.

\begin{figure}
\begin{center}
    \includegraphics[width=10cm]{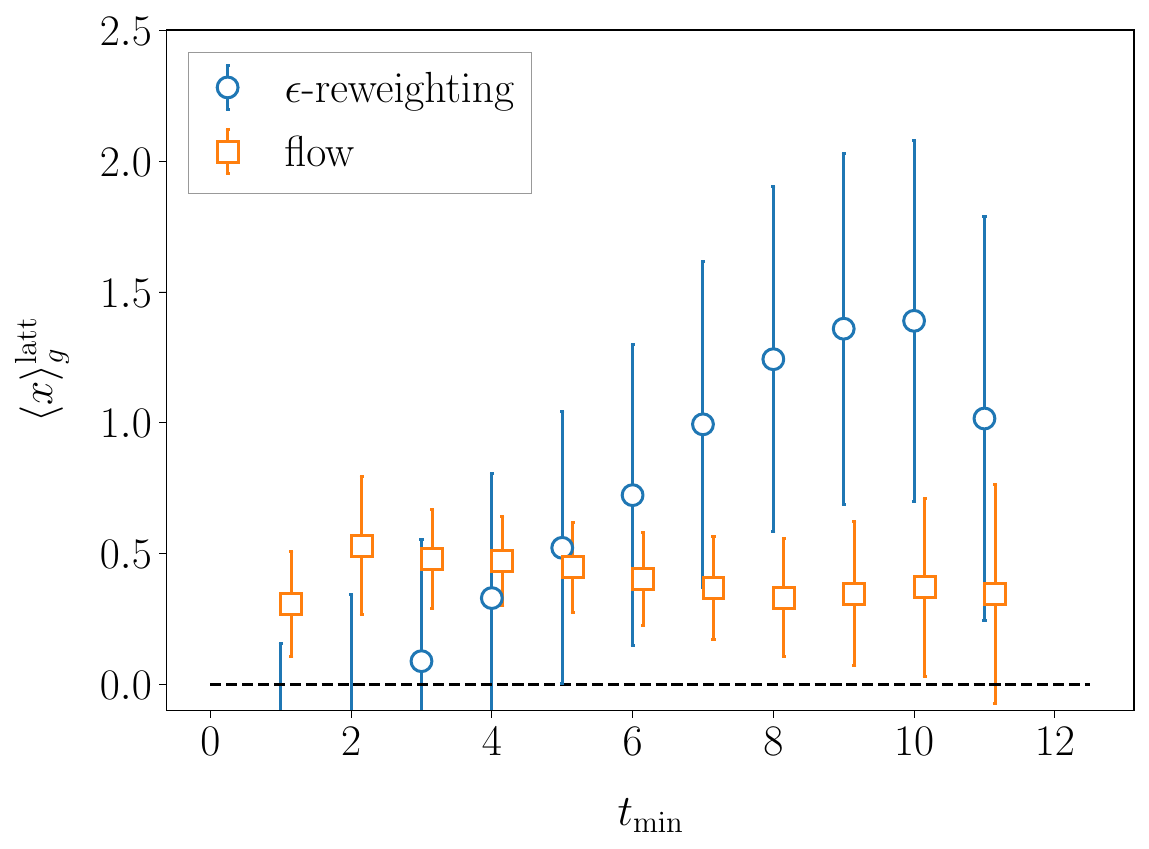}
\end{center}
    \caption{Gluon momentum fraction of the pion computed using the Feynman-Hellmann approach as function of the upper end of the fit range, $t_{\rm min}$. Orange squares correspond to using flows, while blue circles are the baseline of $\epsilon$ reweighting. \label{fig:momfrac}}
\end{figure}

The results are shown in Fig.~\ref{fig:momfrac}. Specifically, we show the extracted gluon momentum fractions of the pion extracted from the fit results for the pion mass~\footnote{16 stochastic sources at random timeslices per configuration are used.} as a function of the fit range, defined as $[t_{\rm min}, T/2]$. We compare the flow approach to $\epsilon$ reweighting. We find that statistical errors are 2-3 times smaller when using flows. This means that for a fixed target statistical uncertainty, 5-10 times fewer configurations are needed.

A final comment about computational costs is due, made with the caveat that the numbers necessarily depend on the implementation and hardware. We present here the costs of this particular demonstration on A100 NVIDIA GPUs. A breakdown of the costs for each approach is:
\begin{align}
\begin{split}
       {\rm COST}(\epsilon{\rm-rew}) &= {\rm generation} + {\rm measurement}, \\
    {\rm COST}({\rm flows}) &= {\rm generation} + {\rm training} + {\rm flow \ eval.} + 2\times {\rm measurement}  + {\rm reweighting} .
\end{split}
\end{align}
The difference for the flow approach is thus that 1) flows need to be trained and evaluated, 2) the reweighting factor involves fermions in the flow case and thus requires inversions to stochastically estimate the determinant ratio,
and 3) an additional measurement is needed on the flowed configurations. We find that both measurement and generation costs per configuration are $0.17 \ {\rm GPU}\cdot{\rm hour}$. In contrast flow evaluation costs barely accounts for $0.01 \ {\rm GPU}\cdot{\rm hour}$, while the reweighting cost per flowed configuration is $0.04 \ {\rm GPU}\cdot{\rm hour}$. Thus, per configuration, the flow approach costs around $2\times$ times more. As for training, this model was trained for 8 days on 16 A100s, and thus 4608 ${\rm GPU}\cdot{\rm hour}$. Given the $5-10\times$ reduction of variance demonstrated in Fig.~\ref{fig:momfrac}, a computational advantage is achieved after 4k configurations if training costs are taken into account. Note that this demonstration used 10k, and thus the results are well into the regime of computational advantage when flows are used.

\section{Conclusion}
Normalizing flows are a promising avenue for accelerating lattice field theory calculations. This work has reviewed two recent advances in applying normalizing flows to 4d gauge theories. The first, learned active loops, provides a substantive improvement to spectral flow models and shows that there are still many possibilities for improving model performance. In the future this work could be extended, for instance by increasing the expressivity of the staple networks using L-CNNs~\cite{Favoni:2020reg}, or by finding other methods to increase the ability of the flow to modify different degrees of freedom. The second advancement presented here is the application of correlated ensemble methods to a theory with dynamical fermions, yielding a true computational advantage in a scenario not far from practical applicability.
Future directions include applying these ideas to larger volumes and lower pion masses, approaching the scale of modern calculations. Separately, the method may be applied for other operators to obtain e.g.~quark mass derivatives and sigma terms.
These two advances are among many improvements in the application of normalizing flows to lattice field theory, bringing this method closer to practical applicability at the state of the art.

\section*{Acknowledgements}
We thank Michael Albergo, Aleksandar Botev, Kyle Cranmer, Jacob Finkenrath, Alexander G.~D.~G.\ Matthews, S\'ebastien Racani\`ere, Ali Razavi, and Danilo J. Rezende for useful discussions and valuable contributions to the early stages of this work. RA, DCH, FRL, PES, and JMU are supported in part by the U.S.\ Department of Energy, Office of Science, Office of Nuclear Physics, under grant Contract Number DE-SC0011090. PES is additionally supported by the U.S.\ DOE Early Career Award DE-SC0021006, by a NEC research award, and by the Carl G and Shirley Sontheimer Research Fund. FRL acknowledges support by the Mauricio and Carlota Botton Fellowship. RA was also partially supported by the High Energy Physics Computing Traineeship for Lattice Gauge Theory (DE-SC0024053). GK was supported by the Swiss National Science Foundation (SNSF) under grant 200020\_200424. This manuscript has been authored by Fermi Forward Discovery Group, LLC under Contract No. 89243024CSC000002 with the U.S. Department of Energy, Office of Science, Office of High Energy Physics. This work is supported by the U.S.\ National Science Foundation under Cooperative Agreement PHY-2019786 (The NSF AI Institute for Artificial Intelligence and Fundamental Interactions, \url{http://iaifi.org/}) and is associated with an ALCF Aurora Early Science Program project, and used resources of the Argonne Leadership Computing Facility which is a DOE Office of Science User Facility supported under Contract DEAC02-06CH11357. The authors acknowledge the MIT SuperCloud and Lincoln Laboratory Supercomputing Center~\cite{reuther2018interactive} for providing HPC resources that have contributed to the research results reported within this paper. Numerical experiments and data analysis used PyTorch~\cite{NEURIPS2019_9015}, JAX~\cite{jax2018github}, Haiku~\cite{haiku2020github}, Horovod~\cite{sergeev2018horovod}, NumPy~\cite{harris2020array}, and SciPy~\cite{2020SciPy-NMeth}. Figures were produced using matplotlib~\cite{Hunter:2007}.

\bibliographystyle{JHEP}
\bibliography{main}

\end{document}